# Completion Time Minimization in Wireless-Powered UAV-Assisted Data Collection System
## (*Full Version*)

Zhen Wang[1,2], Guopeng Zhang[1,2*], Qiu Wang[1,2], Kezhi Wang[3] and Kun Yang[4]

1. The Engineering Research Center of Mine Digitalization of Ministry of Education, China University of Mining and Technology, Xuzhou, China, 221116.
2. The School of Computer Science and Technology, China University of Mining and Technology, Xuzhou, China, 221116.
3. The Department of Computer and Information Science, Northumbria University, Newcastle NE2 1XE, UK.
4. The School of Computer Science and Electronic Engineering, University of Essex, Wivenhoe Park, Colchester, CO4 3SQ, UK.
*\* Correspondence author: gpzhang@cumt.edu.cn*

*Abstract* —In unmanned aerial vehicle (UAV)-assisted data collection system, UAVs can be deployed to charge ground terminals (GTs) via wireless power transfer (WPT) and collect data from them via wireless information transmission (WIT). In this paper, we aim to minimize the time required by a UAV via jointly optimizing the trajectory of the UAV and the transmission scheduling for all the GTs. This problem is formulated as a mixed integer nonlinear programming (MINLP) which are difficult to address in general. To this end, we develop an iterative algorithm based on binary search and successive convex optimization (SCO) to solve it. The simulation shows that our proposed solution outperforms the benchmark algorithms.

## I. Introduction

In some Internet of things (IoT) applications, such as smart farms and opencast coal mine, the environment is normally very large. Lacking of power supply and communication infrastructures make IoT sensors / ground terminals (GTs) difficult to transmit their monitoring data to the control center. Some GTs may rely on wireless multi-hop relay system which may dramatically increase both transmission delay and energy consumption. Recently, UAVs have been applied to support these applications due to the attractive features of high mobility and low operating cost [1]. By installing data receiving module on a UAV, it can act as a mobile base station (BS) to collect the data from GTs by single-hop communication in short range. Moreover, with the help of wireless charging modules, UAV can compensate the energy consumption of GTs by using wireless power transfer (WPT) technology [2]. The authors in [3] optimized the trajectory of a UAV to equitably replenish the energy for a group of GTs but without considering the WIT issue. The authors in [4] reduced the energy consumption of GTs for data uploading by optimizing the UAV trajectory but without considering the UAV-assisted WPT. The authors in [5] divided the operation time of a UAV into WPT and WIT parts. They studied the optimal proportion of the two parts of time but without considering the transmission scheduling problem for GTs. The authors in [6] minimized the outage probability of data transmission in the UAV-assisted WIT and WPT systems. However, they only considered a UAV hovering around the center point of a circular area without optimizing the UAV trajectory. To balance the energy consumption of a group of GTs for data transmission, the authors in [7] jointly optimized hovering position and time of a UAV, but they only focused on the hovering time. The authors in [8] minimized the energy consumption of a UAV for performing wireless-powered data collection. A full-duplex method and a half-duplex method for coordinating the WPT and WIT links are proposed. However, the UAV operated in a fly-then-hover mode, that is, the UAV does not perform any operations during flight, but only performs WPT and WIT during hovering.

In comparison to the above works, the main contributions of our paper are as follows:
1) Instead of considering minimizing the energy consumption for WPT and WIT separately, in this paper, we consider the working/serving time minimization of the UAV, as this affects the mechanical energy consumption, which is considered to be the most important part of the UAV.
2) Moreover, we apply the practical non-linear energy harvesting model [9] for the WPT operation, and a Rician fading channel model [10] for both the WPT and WIT operation. These considerations introduce several non-convex constraints that are difficult to deal with. We present an efficient algorithm to address it successfully.

The remainder of this paper is organized as follows. In Sec. II, we describe the system model. In Sec. III, we present the optimization problem. The algorithm to solve this problem is developed in Sec. IV. In Sec. V, we validate the performance of the algorithm. Finally, we summarize the paper in Sec. VI.

## II. System Model

As shown in Fig. 1, the considered data collection system including a group of fixed GTs, denoted by $\mathcal{M} \triangleq \{1,\cdots,M\}$. A UAV is dispatched to collect the data generated by the GTs, while wireless charging the GTs via WPT technologies. Then, the GTs harvest the energy and use it to upload their information to the UAV.

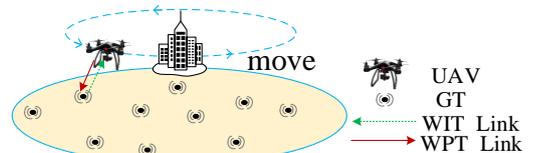

Fig. 1. UAV-assisted WPT and WIT system.

We consider the Cartesian coordinate system where the coordinate of the $m^{\text{th}}$ GT, $\forall m \in \mathcal{M}$ is given by $\mathbf{s}_m = (x_m, y_m, 0)$, which is assumed to be known as a priori for planning the UAV trajectory. We also consider a discrete time model and let $\tau$ denote the duration of a time slot. The location of the UAV is assumed to be approximately unchanged within each slot, and the UAV consumes $T$ consecutive slots to collect the data of the GTs. Let $\mathcal{T} \triangleq \{1, \cdots T\}$ denote the set of the slots. During any $t^{\text{th}}$ slot, $\forall t \in \mathcal{T}$, the UAV flies at a fixed altitude $H$ ($H > 0$) above the ground, and the coordinate of the UAV is given by $\mathbf{q}[t] = (x[t], y[t], H)$.

In practical applications, a UAV starts from an initial point and will return to this point after data collection, as shown in Fig. 1. Denote the initial/final location of the UAV by $\mathbf{q}^{\text{ini}} = (x^{\text{ini}}, y^{\text{ini}}, H)$. Then, one has the following constraint as:

$$\mathbf{q}[1] = \mathbf{q}^{\text{ini}} \text{ and } \mathbf{q}[T] = \mathbf{q}^{\text{ini}}. \tag{1}$$

Let $v^{\max}$ (in meters per second) denote the maximum flight speed of the UAV. The maximum horizontal distance that the UAV can travel in each slot is given by $D^{\max} \triangleq v^{\max}\tau$. During any $t^{\text{th}}$ slot, the following maximum speed constraint for the UAV needs to be satisfied as:

$$\|\mathbf{q}[t+1] - \mathbf{q}[t]\|^2 \leq (D^{\max})^2, \forall t \in \mathcal{T}. \tag{2}$$

In any $t^{\text{th}}$ slot, the channel coefficient between the UAV and the $m^{\text{th}}$ GT can be modeled as

$$h_m[t] = \sqrt{\beta_m[t]}\, g_m[t], \ \forall t \in \mathcal{T}. \tag{3}$$

where $\beta_m[t]$ is the large-scale average channel gain and $g_m[t]$ is the small-scale fading coefficient. Let $d_m[t] = \sqrt{\|\mathbf{q}[t] - \mathbf{s}_m\|^2}$ denote the distance between the UAV and the $m^{\text{th}}$ GT in the $t^{\text{th}}$ slot. Then $\beta_m[t]$ is given as $\beta_m[t] = \rho_0(d_m[t])^{-a}$, where $a$ is the path-loss exponent and $\rho_0$ is the reference channel gain at $d = 1$ m. Since there may exist the line of sight (LoS) links between UAVs and GTs, the small-scale fading $g_m[t]$ can be modeled as Rician fading [10]

$$g_m[t] = \sqrt{\frac{k_m[t]}{k_m[t]+1}}\, g + \sqrt{\frac{1}{k_m[t]+1}}\, \tilde{g}, \forall t \in \mathcal{T}. \tag{4}$$

where $g$ corresponds to the LoS component, $\tilde{g}$ corresponds to the Rayleigh fading component, and $k_m[t]$ is the Rician factor which is the ratio of power of the LoS component to that of the fading component. It is shown in [10] that $k_m[t]$ is closely related to the elevation angle between a UAV and a GT. The expression for $k_m[t]$ is given by $k_m[t] = A_1 \exp(A_2 \theta_m[t])$, where $A_1$ and $A_2$ are constants determined by the environment, and $\theta_m[t] = \arcsin(H/d_m[t])$ is the elevation angle between the UAV and the $m^{\text{th}}$ GT during the $t^{\text{th}}$ slot.

Let $B$ (in Hertz) denote the channel bandwidth of the system and $\sigma^2$ denote the variance of the Gaussian noise at the UAV. Let $\Gamma$ denote the signal-to-noise ratio (SNR) gap. We represent the outage probability that the UAV cannot successfully receive the transmitted data from a GT by $\varepsilon_0$. Let $P^{\text{D}}$ denote the transmission power applied by a GT for data uploading. In order to ensure a high successful probability, e.g., $0 < \varepsilon_0 \leq 0.1$, the maximum transmission rate of the $m^{\text{th}}$ GT during the $t^{\text{th}}$ slot is chosen as [10]

$$R_m[t] = B\log_2\left(1 + \left(C_1 + \frac{C_2}{1+\exp(-(B_1 + B_2 v_m[t]))}\right) \frac{\gamma}{(d_m[t])^a}\right). \tag{5}$$

where $B_1 < 0$, $B_2 > 0$, $C_1 > 0$ and $C_2 > 0$ are constants, $C_1 + C_2 = 1$, $v_m[t] \triangleq \sin(\theta_m[t])$, and $\gamma = P^{\text{D}}\rho_0/\sigma^2\Gamma$. The derivation process of obtaining eq. (5) is detailed in Appendix A.

## III. PROBLEM FORMULATION

Let $L_m$ denote the number of information (in bits) that the $m^{\text{th}}$ GT needs to upload to the UAV. Our objective is to minimize the mission completion time $T$ for the UAV to collect the data of all the GTs. The harvest-then-transmit protocol [2] is adopted in the system and the $t^{\text{th}}$ slot is divided into $M + 1$ non-overlapped intervals as shown in Fig. 2.

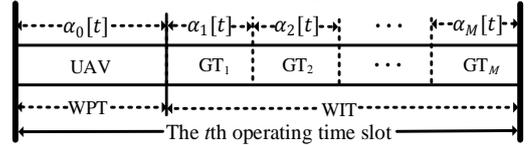

Fig. 2. The structure of the system.

For any $t^{\text{th}}$ slot, the first part $\alpha_0[t]$ ($0 \leq \alpha_0[t] \leq 1$) is allocated to the UAV to deliver RF energy to the GTs by broadcasting, while the remaining $1 - \alpha_0[t]$ part is allocated to the GTs for data uploading. Due to the broadcast nature of WPT, all GTs can receive the energy transmitted by the UAV at the same time. Therefore, $\alpha_0$ is not divided for the GTs. In order to avoid co-channel interference, the GTs are scheduled to upload their data bits orthogonally. Assuming that the $m^{\text{th}}$ GT is allocated $\alpha_m[t]$ part of a slot for data transmission, the following constraints should be satisfied.

$$0 \leq \alpha_m[t] \leq 1, \forall m \in \mathcal{M} \cup \{0\} \text{ and } \sum_{m=0}^{M} \alpha_m[t] \leq 1. \tag{6}$$

Let $P^{\text{U}}$ denote the transmission power of the UAV. After channel attenuation, the input power at the energy harvesting (EH) module of the $m^{\text{th}}$ GT is given by

$$P_m^{\text{in}}[t] = \left(C_1 + \frac{C_2}{1+\exp(-(B_1 + B_2 v_m[t]))}\right) \frac{\delta}{(d_m[t])^a}. \tag{7}$$

where $\delta = P^{\text{U}}\rho_0$. In order to capture the dynamics of the RF energy conversion efficiency for different input power levels, this paper considers a non-linear energy harvesting (EH) model presented in [9]. By using this model, the output power at the energy receiver of the $m^{\text{th}}$ GT is given by

$$\Phi_m = \frac{\Psi_m - K\Omega}{1 - \Omega}. \quad \forall m \in \mathcal{M} \tag{8}$$

where K, $\beta_1$ and $\beta_2$ are constants related to the circuit specifications, $\Psi_m = \frac{K}{1+\exp(-\beta_1(P_m^{\text{in}}-\beta_2))}$, and $\Omega = \frac{1}{1+\exp(\beta_1\beta_2)}$.

By using eq. (8), the amount of energy harvested by the $m^{\text{th}}$ GT during the $t^{\text{th}}$ slot is given by

$$E_m^{\text{H}}[t] = \tau\alpha_0[t]\Phi_m = \tau\alpha_0[t]\left(\frac{D_1}{1+\exp(-\beta_1(P_m^{\text{in}}[t]-\beta_2))} - D_2\right) \tag{9}$$

where $D_1 = \text{K}/(1-\Omega)$ and $D_2 = \text{K}\Omega/(1-\Omega)$.

Since the energy for data transmission is harvested from the UAV, the power control for the GTs is not considered for simplicity. We simply assume a common transmission power $P^{\text{D}}$ for each GT. During the $t^{\text{th}}$ slot, the energy consumed by the $m^{\text{th}}$ GT for data transmission is given as

$$E_m^{\text{C}}[t] = \tau P^{\text{D}}\alpha_m[t], \forall t \in \mathcal{T} \text{ and } \forall m \in \mathcal{M}. \tag{10}$$

According to [2], the following causality constraint on the power management can be considered,

$$E_m^{\text{R}}[t] + E_m^{\text{H}}[t] \geq E_m^{\text{C}}[t], \forall t \in \mathcal{T} \text{ and } \forall m \in \mathcal{M}, \tag{11}$$

where $E_m^{\text{R}}[t]$ denotes the remaining energy of the $m^{\text{th}}$ GT at the beginning of the $t^{\text{th}}$ slot. It is assumed that

$$E_m^{\text{R}}[1] = 0, \forall m \in \mathcal{M}. \tag{12}$$

Constraint (11) states that the energy that has not been accumulated yet cannot be used at the current time slot.

Finally, we also have the following data integrity constraint for the GTs as:

$$\sum_{t=1}^{T} \tau\alpha_m[t]R_m[t] \geq L_m, \ \forall m \in \mathcal{M}. \tag{13}$$

Let $\mathbf{A} = (\alpha_m[t])_{T \times (M+1)}$ and $\mathbf{Q} = (\mathbf{q}[t])_{T \times 1}$. Under the constraints above, the optimization problem to minimize the mission completion time $T$ can be formulated as

$$\min_{\mathbf{A},\mathbf{Q}} T \tag{14}$$

$$\text{s.t. } (1), (2), (6), (11), (12) \text{ and } (13) \tag{14.1}$$

As constraints (11) and (13) are non-convex with respect to $\mathbf{Q}$, problem (14) is challenging to solve. There is no standard method to solve such a Mixed Integer Nonlinear Programming (MINLP) problem efficiently.

## IV. UAV Trajectory Optimization

In this section, we present an effective algorithm to solve problem (14). The main algorithms are as follows.

1) With any given task completion time $T$, one can minimize the difference between the collected data and the target one, which is given as

$$\min_{\mathbf{Q},\mathbf{A},\theta} \theta \tag{15}$$

$$\text{s.t. } (1), (2), (6), (11), (12) \tag{15.1}$$

$$\sum_{t=1}^{T} \alpha_m[t]\, \tau\, R_m[t] \geq L_m - \theta, \forall m \in \mathcal{M} \tag{15.2}$$

2) After obtaining the solution of problem (15), one can determine the next operation according to the value of $\theta$. If $\theta \leq 0$, the current $T$ is feasible to problem (14), otherwise, one has to find a new feasible solution by using binary search algorithm (BSA).

We note that constraints (11) and (15.2) are still non-convex with respect to $\mathbf{Q}$. To make the problem tractable, we divide problem (15) into two sub-problems. In the first sub-problem (called GT scheduling problem or GSP), the scheduling strategy $\mathbf{A}$ for the GTs is optimized with a given UAV trajectory $\mathbf{Q}$. In the second sub-problem (called the UAV trajectory planning problem or UTP), the UAV trajectory $\mathbf{Q}$ is optimized with the obtained $\mathbf{A}$. The GSP and UTP are solved iteratively to find the optimal solution to problem (15).

### A. Solving the GSP

With any given UAV trajectory $\mathbf{Q}$, the GSP is derived from problem (15) as

$$\min_{\mathbf{A},\theta} \theta \tag{16}$$

$$\text{s.t. } (1), (2), (6), (11), (12), \text{ and } (15.2) \tag{16.1}$$

Problem (16) is a standard linear programing (LP), which can be readily solved by applying classical optimization methods.

### B. Solving the UTP

Once the GSP is solved, the optimal $\mathbf{A}$ can be obtained. The UTP is then derived from problem (15) which is given as

$$\min_{\mathbf{Q},\theta} \theta \tag{17}$$

$$\text{s.t. } (1), (2), (11), (12), \text{ and } (15.2) \tag{17.1}$$

However, constraints (11) and (15.2) are still non-convex with respect to $\mathbf{Q}$. Next, we implement an SCO-based algorithm to find the approximate solution of the problem. The main method is to convert constraints (11) and (15.2) into convex problem. Then, we can get the approximate solution to problem (17) by solving a series of convex optimizations iteratively.

Let $\mu$ denote the number of the iterations. In the $\mu^{th}$ iteration, we denote the location of the UAV in the $t^{th}$ slot by $\mathbf{q}^\mu[t]$, and denote the trajectory of the UAV by $\mathbf{Q}^\mu = \{\mathbf{q}^\mu[t], \forall t \in \mathcal{T}\}$.

It is well known that the first-order Taylor expansion of any convex function at any point is its global lower bound. By inspecting eq. (5), we note that $R_m[t]$ is convex with respect to $\exp(-(B_1 + B_2 v_m[t])) - \exp(-(B_1 + B_2 v_m^\mu[t]))$ and $\|\mathbf{q}[t] - \mathbf{s}_m\|^2 - \|\mathbf{q}^\mu[t] - \mathbf{s}_m\|^2$. Then, one can get the following Lemma 1.

**Lemma 1**: Denote the lower bound of $R_m[t]$ in the $(\mu + 1)^{th}$ iteration by $R_m^{lb}[t]$, which is given as

$$R_m[t] \geq R_m^{lb}[t] \triangleq R_m^\mu[t] - X_m^\mu[t]\left(e^{-S_m[t]} - e^{-S_m^\mu[t]}\right) - Y_m^\mu[t](\|\mathbf{q}[t] - \mathbf{s}_m\|^2 - \|\mathbf{q}^\mu[t] - \mathbf{s}_m\|^2) \tag{18}$$

where

$$R_m^\mu[t] = B\log_2\left(1 + \left(C_1 + \frac{C_2}{1+e^{-S_m^\mu[t]}}\right)\frac{\gamma}{(d_m^\mu[t])^\alpha}\right) \tag{18.1}$$

$$X_m^\mu[t] = \frac{\gamma C_2 B}{\ln 2\left(1+e^{-S_m^\mu[t]}\right)\left(\gamma\left(C_1\left(1+e^{-S_m^\mu[t]}\right)+C_2\right)+\left(1+e^{-S_m^\mu[t]}\right)(d_m^\mu[t])^\alpha\right)} \tag{18.2}$$

$$Y_m^\mu[t] = \frac{\gamma \alpha B\left(C_1\left(1+e^{-(B_1+B_2 v_m^\mu[t])}\right)+C_2\right)}{\ln 4(d_m^\mu[t])^2\left(\gamma\left(C_1\left(1+e^{-S_m^\mu[t]}\right)+C_2\right)+\left(1+e^{-S_m^\mu[t]}\right)(d_m^\mu[t])^\alpha\right)} \tag{18.3}$$

$$S_m[t] = B_1 + B_2 v_m[t] \tag{18.4}$$

$$S_m^\mu[t] = B_1 + B_2 v_m^\mu[t] \tag{18.5}$$

*Proof*: Please refer to Appendix B.

By checking eq. (18), we note that $R_m^{lb}[t]$ is concave with respect to $\mathbf{q}[t]$ and $S_m[t]$, and constraint (18.4) is tight which limits the feasible domain of the problem. Similar to [10], one can slack the equality constraint (18.4) as

$$S_m[t] \leq B_1 + B_2 v_m[t] \tag{19}$$

Constraint (19) is non-convex with respect to $\mathbf{q}[t]$. To address this problem, we give the following Lemma 2, which can convert constraint (19) into a convex one with respect to $\mathbf{q}[t]$.

**Lemma 2:** Constraint (19) can be converted into a convex constraint with respect to $\mathbf{q}[t]$ as

$$S_m[t] \leq B_1 + B_2 v_m^{lb}[t] \tag{20}$$

where $v_m^{lb}[t] \triangleq v_m^\mu[t] - \frac{H}{2(\|\mathbf{q}^\mu[t]-\mathbf{s}_m\|^2)^{3/2}} \times (\|\mathbf{q}[t] - \mathbf{s}_m\|^2 - \|\mathbf{q}^\mu[t] - \mathbf{s}_m\|^2)$ is the definite lower bound for $v_m[t]$.

*Proof*: Please refer to Appendix C.

Now, we can convert constraint (15.2) into a convex constraint with respect to $\mathbf{q}[t]$ as

$$\sum_{t=1}^{T} \alpha_m[t]\tau R_m^{lb}[t] \geq L_m - \theta, \forall m \in \mathcal{M} \tag{21}$$

Next, for eq. (11), one can see that $E_m^H[t]$ is convex with respect to $\exp\left(-\beta_1(P_m^{in}[t]-\beta_2)\right) - \exp\left(-\beta_1(P_m^{in,\mu}[t]-\beta_2)\right)$. Thus, we have the following Lemma 3.

**Lemma 3:** There exists a definite lower bound $E_m^{H,lb}[t]$ for $E_m^H[t]$ in the $(\mu+1)^{th}$ iteration, which is given as

$$E_m^H[t] \geq E_m^{H,lb}[t] \triangleq E_m^{H,\mu}[t] - \chi_m^\mu[t]\left(e^{-U_m[t]} - e^{-U_m^\mu[t]}\right) \tag{22}$$

$$E_m^{H,\mu}[t] = \tau\alpha_0[t]\left(\frac{D_1}{1+\exp\left(-\beta_1\left(P_m^{in,\mu}[t]-\beta_2\right)\right)} - D_2\right) \tag{22.1}$$

$$\chi_m^\mu[t] = \frac{\tau\alpha_0[t]D_1}{\left(1+e^{-U_m^\mu[t]}\right)^2} \tag{22.2}$$

$$U_m[t] = \beta_1 P_m^{in} - \beta_1\beta_2 \tag{22.3}$$

$$U_m^\mu[t] = \beta_1 P_m^{in,\mu}[t] - \beta_1\beta_2 \tag{22.4}$$

*Proof*: Please refer to Appendix D.

By analyzing eq. (22.3), we note that $E_m^{H,lb}[t]$ is concave with respect to $\mathbf{q}[t]$ and $U_m[t]$. Since the equality constraint (22.3) is tight for the problem to find a feasible solution, we relax it as

$$U_m[t] \leq \beta_1 P_m^{in}[t] - \beta_1\beta_2 . \tag{23}$$

The added constraint (23) is non-convex with respect to $\mathbf{q}[t]$. To address this problem, we give the following Lemma 4, which converts constraint (23) into a convex one.

**Lemma 4:** The non-convex constraint in (23) can be converted into a convex one as

$$U_m[t] \leq \beta_1 P_m^{\text{in,lb}}[t] - \beta_1 \beta_2, \quad (24)$$

where

$$P_m^{\text{in,lb}}[t] \triangleq P_m^{\text{in},\mu}[t] - \psi_m^{\mu}[t]\left(e^{-S_m[t]} - e^{-S_m^{\mu}[t]}\right) -$$
$$\eta_m^{\mu}[t](\|\mathbf{q}[t] - \mathbf{s}_m\|^2 - \|\mathbf{q}^{\mu}[t] - \mathbf{s}_m\|^2) \quad (24.1)$$

$$P_m^{\text{in},\mu}[t] = \left(C_1 + \frac{C_2}{1+\exp(-S_m^{\mu}[t])}\right)\frac{\delta}{(d_m^{\mu}[t])^a} \quad (24.2)$$

$$\psi_m^{\mu}[t] = \frac{C_2 \delta}{\left(1+e^{-S_m^{\mu}[t]}\right)^2 (d_m^{\mu}[t])^a} \quad (24.3)$$

$$\eta_m^{\mu}[t] = \frac{(a/2)\delta\left(C_1\left(1+e^{-S_m^{\mu}[t]}\right)+C_2\right)}{\left(1+e^{-S_m^{\mu}[t]}\right)(d_m^{\mu}[t]^2)^{a/2+1}} \quad (24.4)$$

*Proof*: Please refer to Appendix E.

Now, constraint (11) is also converted into a convex one with respect to $\mathbf{q}[t]$ as

$$E_m^{\text{R}}[t] + E_m^{\text{H,lb}}[t] \geq E_m^{\text{C}}[t], \forall t \in \mathcal{T} \text{ and } \forall m \in \mathcal{M}. \quad (25)$$

Based on the above analysis, we can approximate problem (17) into the following convex problem with respect to $\mathbf{q}[t]$ as

$$\min_{\mathbf{Q},S,U,\theta} \theta \quad (26)$$

$$\text{s.t. } (1),(2),(12),(20),(21),(24), \text{ and } (25) \quad (26.1)$$

which can be solved by using standard convex optimization solvers such as CVX [11].

### C. Overall Algorithm

With any given task completion time $T$, one can solve problem (16) and problem (26) iteratively to find the solution to problem (15). The initial trajectory of the UAV is represented as $\mathbf{Q}^{\mu=0} = \{\mathbf{q}^{\mu=0}[t], \forall t \in \mathcal{T}\}$. The method of finding the feasible $\mathbf{Q}^{\mu=0}$ is detailed in Appendix F.

Based on the above analysis, the developed algorithm is shown in **Algorithm 1**.

**Algorithm 1** Iterative algorithm for solving problem (15)

1. **Initialize:** Let $\mu = 0$. Initializes $\mathbf{Q}^0$ as the trajectory of the UAV using the method proposed in Appendix F.
2. **Repeat:**
3. For given UAV trajectory $\mathbf{Q}^{\mu}$, obtain scheduling $\mathbf{A}^{\mu+1}$ by solving problem (16).
4. For given $\mathbf{A}^{\mu+1}$, obtain the UAV trajectory $\mathbf{Q}^{\mu+1}$ by solving problem (26).
5. Update $\mu = \mu + 1$.
6. **Until:** The decrease of the objective $\theta$ is below a predefined threshold $\varphi > 0$ or a maximum number of iterations ($\mu^{\max}$) is reached.
7. **Return:** The scheduling matrix $\mathbf{A}$, the UAV trajectory $\mathbf{Q}$, and the value of $\theta$.

In the simulations, we find that the algorithm converges to the stable point after about 20 iterations by using the initial trajectory selection algorithm presented in Appendix F. After executing Algorithm 1, if the obtained objective value $\theta \leq 0$, it means that the current $T$ is feasible to problem (14). In such a case, one has to find a new feasible solution by using binary search algorithm (BSA). The overall BSA for finding the approximate solution to problem (14) is summarized in **Algorithm 2**.

In most scenario such as environment detection, the location of GTs is normally fixed, and, due to the periodic execution of perceptual tasks, the amount of data generated by GTs can be predicted. Therefore, the algorithm can be performed offline at the cloud. The obtained UAV flight trajectory, WPT and data collection scheduling are imported into the UAV before starting the mission. This avoids the operational data exchange between the UAV and the cloud during mission execution, and can reduce the computational capacity required in UAV.

**Algorithm 2** The BSA to solve problem (14)

1. **Initialize:** Let $T^{\min}$ and $T^{\max}$ be the lower and upper bounds of the BSA to find the optimal $T$, respectively.
2. **Repeat:**
3. Update $T = \left\lceil \frac{T^{\min}+T^{\max}}{2} \right\rceil$.
4. For given $T$, obtain the intermediate results of $\mathbf{A}$, $\mathbf{Q}$, and $\theta$ by performing **Algorithm 1**.
5. If $\theta \leq 0$, set $T^{\max} = T$.
6. Else, set $T^{\min} = T$.
7. **Until:** $(T^{\max} - T^{\min}) \leq 1$.
8. Let $T = T^{\min}$.
9. Obtain the intermediate results of $\mathbf{A}$, $\mathbf{Q}$, and $\theta$ by performing **Algorithm 1**.
10. If $\theta \leq 0$, the optimal solution to problem (14) is determined as $\mathbf{A}^* = \mathbf{A}$, $\mathbf{Q}^* = \mathbf{Q}$ and $T^* = T$.
11. Else, let $T = T^{\max}$.
12. The optimal solution to problem (14), i.e., $\mathbf{A}^*$, $\mathbf{Q}^*$ and $T^*$ can be obtained by performing **Algorithm 1**.

## V. NUMERICAL SIMULATION

The available time of a UAV is constrained by its mechanical energy consumption and portable energy. So, in the considered single-UAV supported data collection system, the proposed solution can provide data collection services for a small and medium-sized Internet of things. The GTs are randomly distributed in a square area of $500 \times 500$ m$^2$, and the horizontal flying speed of the UAV is $V_h^{\max} = 35$ m/s. The energy broadcasting power of the UAV is $P^U = 40$ dBm [3]. The transmission power of each GT is $P^D = 20$ dBm. The system bandwidth is $B = 1$ MHz, the path-loss exponent is $a = 2$, the reference channel power gain at $d = 1$ m is $\rho_0 = -10$ dB, the SNR gap is $\Gamma = 8.2$ dB, and the noise power at the receiver of the UAV is $\sigma^2 = 10^{-10}$ W. The rest of the parameters used in the simulation are: $\tau = 1$ s, $\mathbf{q}^{\text{ini}} = (0,0)$, $K = 0.02$, $\beta_1 = 6400$, $\beta_2 = 0.003$, $B_1 = -4.3221$, $B_2 = 6.075$, $C_1 = 0$, and $C_2 = 1$. In the simulation, the following two algorithms are implemented for comparison purpose. The first algorithm, called the *GT Scheduling Algorithm* (GSA), only optimizes the transmission scheduling of the GTs but without optimizing the UAV trajectory. The UAV trajectory is obtained by using the trajectory initialization method. The second algorithm, called the *Collect data when Hovering Algorithm* (CHA), requires the UAV to hover over each GT when performing WPT and data collection. The aim of the CHA is to minimize the total flight time of the UAV. This problem is a typical TSP and is solved by using genetic algorithms.

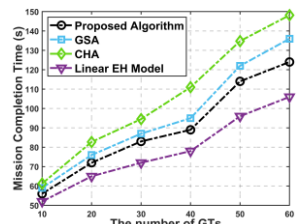 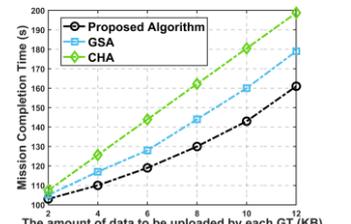

Fig. 3 The completion time vs. the number of the GTs.  Fig. 4 The completion time vs the amount of data of each GT.

First, we set the amount of data of each GT as 5 KB and the altitude of the UAV as 20m. In Fig. 3, we show the trend of the mission completion time with the increase of the number of the GTs from 10 to 60. One notices that most of the existing works on WPT networks, e.g., [3] are based on a linear EH model. In order to show the impact of the non-linear EH model on the system performance, we depict the curve when a linear EH model (50% power conversion efficiency) is used. It is observed that our proposed algorithm outperforms the GSA and CHA, as expected. It also indicates that the linear EH model cannot properly model the power dependent EH efficiency which leads to a mismatch of resource allocation and overestimation of system performance. In Fig. 4, we apply the above settings but fix the number of the GTs to 50. One can see that with the increase of the amount of data to upload, the performance gap between the proposed algorithm and other two algorithms increases, which demonstrates the effectiveness of our solution. It is noted that the UAV synchronously performs WPT and data collection during flight. So the task completion time of the UAV shown in Fig. 3, 4 and 7 just equals to the flight time of the UAV. In addition, as the task completion time of the UAV is optimized under a given flight speed, the shorter the time consumed by the UAV to complete the task, the lower the energy consumption of flight.

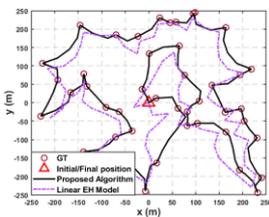
Fig. 5. The trajectory of the UAV.
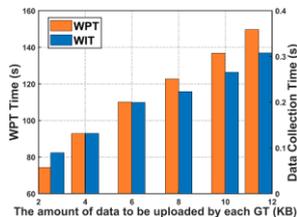
Fig. 6 The time allocation of the UAV.

Next, we set the number of the GTs as 50 and each GT has 5 KB of data to upload. The schematic diagram of the optimized UAV trajectory in shown Fig. 5. It is observed that the proposed algorithm guides the UAV starting from the data center (marked by a red triangle), and then return to the starting point after completing the task. Moreover, the trajectory obtained by using the nonlinear EH model makes the UAV fly longer than that obtained by using the linear EH model. This is because the nonlinear model requires the UAV to spend more time (that is, flying longer distance) charging the GTs.

In Fig. 6, we show the time allocation of the UAV for data collection (i.e., WIT) and WPT. Note that the data collection time and the WPT time are represented in different vertical coordinates in the figure, respectively. It is observed that with the linear increase of the uploaded data, the required transmission time and energy also increase linearly. However, the power conversion efficiency of the EH module leads to a non-linear increase of the WPT time. This verifies that the proposed algorithm can adaptively adjust the WPT time of the UAV according to the amount of the uploaded data and the power conversion efficiency of the GTs.

In Fig. 7, we show the impact of UAV's flight altitude on the system performance. Here, we fix the number of the GTs as 30 and set the amount of data of each GT as 2 KB.

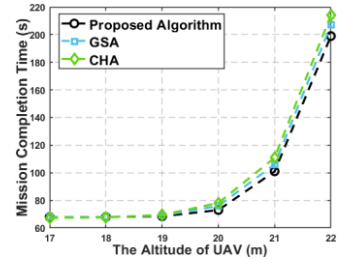
Fig. 7 The mission completion time versus the number of the GTs.

When the flying altitude of the UAV exceeds 20 meters, the mission completion time rises sharply. This is partly caused by the non-linear power conversation efficiency of EH modules. When the UAV flies above a certain altitude, the power received by the GTs is reduced sharply due to channel fading, and the low conversation efficiency at low power input leads to a sharp increase in the EH time. This further leads to a sharp increase in the overall mission completion time of the UAV.

## VI. CONCLUSIONS AND FUTURE WORKS

This paper has studied the joint trajectory optimization and transmission scheduling in a wireless-powered UAV-assisted data collection system. An iterative and efficient algorithm has been developed to minimize the data collection time of the UAV. In the future, we plan to extend our work to multi-UAV scenario, where the issues of GT-and-UAV association, the synchronization and interference control of the UAVs, and the obstacle avoidance and energy constraints for the UAVs will be considered.

## Appendix A

*The derivation process of obtaining the rate expression (5).*

The maximum achievable rate between the UAV and the $m^{\text{th}}$ GT in the $t^{\text{th}}$ time slot is given as

$$C_m[t] = B\log_2\left(1 + \frac{|h_m[t]|^2 P^{\text{D}}}{\sigma^2 \Gamma}\right), \forall t \in \mathcal{T}. \quad (27.1)$$

Given the transmission rate of the $m^{\text{th}}$ GT in the $t^{\text{th}}$ time slot by $R_m[t]$, the transmitted data from the GT cannot be successfully received by UAV if the maximum achievable rate $C_m[t]$ is smaller than the threshold $R_m[t]$. Thus, the outage probability $P_m[t]$ that the UAV cannot successfully receive the transmitted data from the GT can be expressed as

$$P_m[t] = \mathbb{P}(C_m[t] < R_m[t]) = F_{m,t}\left(\frac{\sigma^2\Gamma(2^{R_m[t]}-1)}{\beta_m[t]P^{\text{D}}}\right). \quad (27.2)$$

where $F_{m,t}$ is the non-decreasing cumulative distribution function (CDF) of the random variable $|g_m[t]|^2$. For Rician fading, the CDF of $|g_m[t]|^2$ can be explicitly expressed as

$$F_{m,t}(\varpi) = 1 - Q_1\left(\sqrt{2k_m[t]}, \sqrt{2(k_m[t]+1)\varpi}\right). \quad (27.3)$$

where $Q_1(\varrho, \varsigma)$ is the standard Marcum-Q function.

In order to ensure that the data transmitted by the GT is reliable received by the UAV, $R_m[t]$ is chosen such that $P_m[t] = \varepsilon_0$ ($0 < \varepsilon_0 \leq 0.1$), where $\varepsilon_0$ is the maximum tolerable outage probability. By substituting $\beta_m[t] = \rho_0(d_m[t])^{-a}$ into eq. (27.2) with $P_m[t] = \varepsilon_0$, we get

$$R_m[t] = B\log_2\left(1 + \frac{\delta_m[t]\gamma}{(d_m[t])^a}\right), \forall t \in \mathcal{T}. \quad (27.4)$$

where $\gamma = \frac{P^{\text{D}}\rho_0}{\sigma^2\Gamma}$ and $\delta_m[t]$ is the solution to $F_{m,t}(\varpi) = \varepsilon_0$. Due to the lack of an explicit form for the inverse Marcum-Q function, $\delta_m[t]$ is intractable. Because it is difficult to get a closed-form expression of $\delta_m[t]$, we use the result provided by ref. [10] to approximate $\delta_m[t]$ to the following logistic function

$$\delta_m[t] \approx C_1 + \frac{C_2}{1+\exp(-(B_1+B_2 v_m[t]))}. \quad (27.5)$$

By substituting (27.5) into eq. (27.4), we can finally obtain eq. (5).

## APPENDIX B

*Proof of Theorem 1:* First, we prove that $R_m[t]$ is convex with respect to $1 + \exp(-(B_1 + B_2 v_m[t]))$ and $\|\mathbf{q}[t] - \mathbf{s}_m\|^2$.

To this end, we define $x = 1 + \exp(-(B_1 + B_2 v_m[t]))$ and $y = \|\mathbf{q}[t] - \mathbf{s}_m\|^2$. Let $C_3 = C_1 \gamma > 0$ and $C_4 = C_2 \gamma > 0$. By substituting $x$, $y$, $C_3$ and $C_4$ into the expression of $R_m[t]$ in eq. (5) and ignoring some constant terms, we can rewrite eq. (5) as $w(x,y) = \log_2\left(1 + \left(C_3 + \frac{C_4}{x}\right)\frac{1}{y^{a/2}}\right)$.

By applying the formula of changing base of logarithmic function to $w(x,y)$, we can get $w(x,y) = g(x,y)\log_2(e)$, where $g(x,y) = \ln\left(1 + \left(C_3 + \frac{C_4}{x}\right)\frac{1}{y^{a/2}}\right)$. Obviously, if $g(x,y)$ were convex with respect to $x$ and $y$, $w(x,y)$ does. Next, we show that $g(x,y)$ is convex with respect to $x$ and $y$.

We derive the Hessian of $g(x,y)$ as

$$\nabla^2 g(x,y) = \begin{bmatrix} \frac{\partial^2 g(x,y)}{\partial x^2} & \frac{\partial^2 g(x,y)}{\partial x \partial y} \\ \frac{\partial^2 g(x,y)}{\partial y \partial x} & \frac{\partial^2 g(x,y)}{\partial y^2} \end{bmatrix} \quad (28)$$

where

$$\frac{\partial^2 g(x,y)}{\partial x^2} = \frac{C_4\left(2xy^{\frac{a}{2}} + 2C_3 x + C_4\right)}{x^2\left(xy^{\frac{a}{2}} + C_3 x + C_4\right)^2} \quad (28.1)$$

$$\frac{\partial^2 g(x,y)}{\partial x \partial y} = \frac{\partial^2 g(x,y)}{\partial y \partial x} = \frac{(a/2)C_4 y^{\frac{a}{2}-1}}{\left(xy^{\frac{a}{2}} + C_3 x + C_4\right)^2} \quad (28.2)$$

$$\frac{\partial^2 g(x,y)}{\partial y^2} = \frac{(a/2)(C_3 x + C_4)\left[(1+a/2)xy^{\frac{a}{2}} + C_3 x + C_4\right]}{y^2\left(xy^{\frac{a}{2}} + C_3 x + C_4\right)^2} \quad (28.3)$$

For any $\mathbf{h} = [h_1, h_2]^T$, given $a \geq 2$, $x > 0$ and $y > 0$, one can get

$$\mathbf{h}^T \nabla^2 g(x,y) \mathbf{h} = h_1^2 \left(\frac{C_4\left(2xy^{\frac{a}{2}} + 2C_3 x + C_4\right)}{x^2\left(xy^{\frac{a}{2}} + C_3 x + C_4\right)^2}\right) +$$

$$h_2^2 \left(\frac{\left(\frac{a}{2}\right)(C_3 x + C_4)\left[\left(1+\frac{a}{2}\right)xy^{\frac{a}{2}} + C_3 x + C_4\right]}{y^2\left(xy^{\frac{a}{2}} + C_3 x + C_4\right)^2}\right) + \frac{h_1 h_2 a C_4 y^{\frac{a}{2}-1}}{\left(xy^{\frac{a}{2}} + C_3 x + C_4\right)^2} \geq 0 \quad (28.4)$$

where $\nabla^2 g(x,y)$ is a positive semidefinite matrix. Thus, $g(x,y)$ is convex with respect to $x$ and $y$. It leads to the convexity of $w(x,y)$.

Next, we prove that $R_m[t]$ is convex with respect to $\exp(-(B_1 + B_2 v_m[t])) - \exp(-(B_1 + B_2 v_m^\mu[t]))$ and $\|\mathbf{q}[t] - \mathbf{s}_m\|^2 - \|\mathbf{q}^\mu[t] - \mathbf{s}_m\|^2$.

Let $j = e^{-(B_1+B_2 v_m[t])} - e^{-(B_1+B_2 v_m^\mu[t])}$ and $k = 1 + e^{-(B_1+B_2 v_m^\mu[t])}$. We can rewrite $x = j + k$. Let $u = \|\mathbf{q}[t] - \mathbf{s}_m\|^2 - \|\mathbf{q}^\mu[t] - \mathbf{s}_m\|^2$ and $n = \|\mathbf{q}^\mu[t] - \mathbf{s}_m\|^2$. We can rewrite $y = u + n$. By substituting $x$ and $y$ into the expression of $R_m[t]$ in eq. (5), we can rewrite eq. (5) as $f(j, u) = B\log_2\left(1 + \left(C_1 + \frac{C_2}{j+k}\right)\frac{\gamma}{(u+n)^{a/2}}\right)$.

It is noted that $k$ and $n$ are constants. Since $w(x,y)$ is convex with respect to $x = j + k$ and $y = u + n$, we know that $f(j, u)$ is convex with respect to $j$ and $u$.

Now, we use the first-order Taylor expansion to approximate the convex function $f(j,u)$ at any $j_0$ and $u_0$ as

$$f(j,u) \geq f(j_0, u_0) + f'_j(j_0, u_0)(j - j_0) + f'_u(j_0, u_0)(u - u_0) \quad (28.5)$$

where

$$f'_j(j_0, u_0) = \frac{-\gamma C_2 B}{\ln 2 (j_0+k)(\gamma(C_1(j_0+k)+C_2)+(j_0+k)(u_0+n)^{a/2})} \quad (28.6)$$

$$f'_k(j_0, u_0) = \frac{-\gamma a B(C_1(j_0+k)+C_2)}{\ln 4 (u_0+n)(\gamma(C_1(j_0+k)+C_2)+(j_0+k)(u_0+n)^{a/2})} \quad (28.7)$$

By letting $j_0 = 0$ and $u_0 = 0$, we obtain

$$B\log_2\left(1 + \left(C_1 + \frac{C_2}{j+k}\right)\frac{\gamma}{(u+n)^{a/2}}\right) \geq B\log_2\left(1 + \left(C_1 + \frac{C_2}{k}\right)\frac{\gamma}{n^{\frac{a}{2}}}\right) - \frac{\gamma C_2 B}{\ln 2 k(\gamma(C_1 k+C_2)+kn^{a/2})} j - \frac{\gamma a B(C_1 k+C_2)}{\ln 4 n(\gamma(C_1 k+C_2)+kn^{a/2})} u \quad (28.8)$$

Finally, we substitute $j = e^{-(B_1+B_2 v_m[t])} - e^{-(B_1+B_2 v_m^\mu[t])}$, $k = 1 + e^{-(B_1+B_2 v_m^\mu[t])}$, $u = \|\mathbf{q}[t] - \mathbf{s}_m\|^2 - \|\mathbf{q}^\mu[t] - \mathbf{s}_m\|^2$, and $n = \|\mathbf{q}^\mu[t] - \mathbf{s}_m\|^2$ into inequality (28.8), and, then, get Lemma 1.

## APPENDIX C

*Proof of Theorem 2:* First, we need to prove that $v_m[t]$ is convex with respect to $\|\mathbf{q}[t] - \mathbf{s}_m\|^2 - \|\mathbf{q}^\mu[t] - \mathbf{s}_m\|^2$.

For ease of presentation, we rewrite $v_m[t] = \frac{H}{\sqrt{\|\mathbf{q}[t] - \mathbf{s}_m\|^2}}$ as $r(u) = \frac{H}{\sqrt{u+n}}$. The second derivative of $r(u)$ is given by

$$r''(u) = \frac{3H}{4(u+n)^{5/2}} \quad (29)$$

For $\forall (u+n) > 0$, $r''(u) = \frac{3H}{4(u+n)^{5/2}} \geq 0$. It means that $r(u)$ is convex with respect to $u$.

Now, we use the first-order Taylor expansion to approximate the convex function $r(u)$ at any $u_0$. as

$$r(u) \geq r(u_0) + r'(u_0)(u - u_0) \quad (29.1)$$

By letting $u_0 = 0$, we can obtain

$$\frac{H}{\sqrt{u+n}} \geq \frac{H}{\sqrt{n}} - \frac{Hu}{2n^{3/2}} \quad (29.2)$$

By substituting $u = \|\mathbf{q}[t] - \mathbf{s}_m\|^2 - \|\mathbf{q}^\mu[t] - \mathbf{s}_m\|^2$ and $n = \|\mathbf{q}^\mu[t] - \mathbf{s}_m\|^2$ into inequality (29.2), we can derive that

$$v_m[t] = \frac{H}{\sqrt{\|\mathbf{q}[t]-\mathbf{s}_m\|^2}} \geq \frac{H}{\sqrt{\|\mathbf{q}^\mu[t]-\mathbf{s}_m\|^2}} - \frac{H}{2(\|\mathbf{q}^\mu[t]-\mathbf{s}_m\|^2)^{3/2}} x = v_m^{\text{lb}}[t] \quad (29.3)$$

Finally, by further substituting eq. (29.3) into inequality (19), we can obtain inequality (20). Lemma 2 is then proved.

## APPENDIX D

*Proof of Theorem 3:* We first prove that $E_m^H[t]$ is convex with respect to $\exp\left(-\beta_1(P_m^{\text{in}}[t] - \beta_2)\right) - \exp\left(-\beta_1(P_m^{\text{in},\mu}[t] - \beta_2)\right)$.

Let $p = e^{-(\beta_1 P_m^{\text{in}}[t] - \beta_1 \beta_2)} - e^{-(\beta_1 P_m^{\text{in},\mu}[t] - \beta_1 \beta_2)}$ and $z = 1 + e^{-(\beta_1 P_m^{\text{in},\mu}[t] - \beta_1 \beta_2)}$. For ease of presentation, we rewrite the expression of $E_m^H[t]$ in eq. (9) as $T(p) = \tau \alpha_0[t]\left(\frac{D_1}{p+z} - D_2\right)$. The second derivative of $T(p)$ is given by

$$T''(p) = \frac{2\tau\alpha_0[t]D_1}{(p+z)^3} \tag{30}$$

For $\forall(p+z) > 0$, we have $T''(p) \geq 0$. It means that $T(p)$ is convex with respect to $p$.

Now, we use the first-order Taylor expansion to approximate $T(p)$ at any $p_0$ as

$$T(x) \geq T(p_0) + T'(p_0)(p - p_0) \tag{30.1}$$

By letting $p_0 = 0$, we can obtain

$$E_m^{\text{H}}[t] = \tau\alpha_0[t]\left(\frac{D_1}{p+z} - D_2\right)$$
$$\geq \tau\alpha_0[t]\left(\frac{D_1}{z} - D_2\right) - \frac{\tau\alpha_0[t]D_1}{z^2}x = E_m^{\text{H,lb}}[t] \tag{30.2}$$

By substituting $p = e^{-(\beta_1 P_m^{\text{in}}[t] - \beta_1\beta_2)} - e^{-(\beta_1 P_m^{\text{in},\mu}[t] - \beta_1\beta_2)}$ and $z = 1 + e^{-(\beta_1 P_m^{\text{in},\mu}[t] - \beta_1\beta_2)}$ into inequality (30.2), we can prove Lemma 3.

## APPENDIX E

*Proof of Theorem 4:* First, we prove that $P_m^{\text{in}}[t]$ is convex with respect to $1 + \exp(-(B_1 + B_2 v_m[t]))$ and $\|\mathbf{q}[t] - \mathbf{s}_m\|^2$.

By substituting $x$, $y$, $C_3$ and $C_4$ into the expression of $P_m^{\text{in}}[t]$ in eq. (7) and ignoring some constant terms, we can rewrite eq. (7) as $F(x,y) = \left(C_3 + \frac{C_4}{x}\right)\frac{1}{y^{a/2}}$. Next, we show that $F(x,y)$ is convex with respect to $x$ and $y$.

We derive the Hessian of $F(x,y)$ is

$$\nabla^2 F(x,y) = \begin{bmatrix} \frac{\partial^2 F(x,y)}{\partial x^2} & \frac{\partial^2 F(x,y)}{\partial x \partial y} \\ \frac{\partial^2 F(x,y)}{\partial y \partial x} & \frac{\partial^2 F(x,y)}{\partial y^2} \end{bmatrix} \tag{31}$$

where

$$\frac{\partial^2 F(x,y)}{\partial x^2} = \frac{2C_4}{x^3 y^{\frac{a}{2}}} \tag{31.1}$$

$$\frac{\partial^2 F(x,y)}{\partial x \partial y} = \frac{\partial^2 F(x,y)}{\partial y \partial x} = \frac{aC_4}{2x^2 y^{\frac{a}{2}+1}} \tag{31.2}$$

$$\frac{\partial^2 F(x,y)}{\partial y^2} = \frac{a(\frac{a}{2}+1)(C_3 x + C_4)}{2(xy^{\frac{a}{2}+1})} \tag{31.3}$$

For any $\mathbf{h} = [h_1, h_2]^T$, given $a \geq 2$, $x > 0$ and $y > 0$, we have

$$\mathbf{h}^T \nabla^2 F(x,y) \mathbf{h} = h_1^2\left(\frac{2C_4}{x^3 y^{\frac{a}{2}}}\right) + h_2^2\left(\frac{a(\frac{a}{2}+1)(C_3 x + C_4)}{2(xy^{\frac{a}{2}+1})}\right) +$$

$$\frac{2h_1 h_2 aC_4}{2x^2 y^{\frac{a}{2}+1}} = \frac{2h_1 h_2 aC_4 xy + 4h_1^2 C_4 y^2 + \frac{3}{2}h_2^2 ax^2(C_3 x + C_4)}{2x^3 y^{\frac{5a}{2}}} \geq 0 \tag{31.4}$$

$\nabla^2 F(x,y)$ is a positive semidefinite matrix. Thus, $F(x,y)$ is convex with respect to $x$ and $y$.

Next, we prove that $P_m^{\text{in}}[t]$ is convex with respect to $\exp(-(B_1 + B_2 v_m[t])) - \exp\left(-(B_1 + B_2 v_m^\mu[t])\right)$ and $\|\mathbf{q}[t] - \mathbf{s}_m\|^2 - \|\mathbf{q}^\mu[t] - \mathbf{s}_m\|^2$.

We can rewrite $x = j + k$ and $y = u + n$. By substituting $x$ and $y$ into the expression of $P_m^{\text{in}}[t]$ in eq. (7), we can rewrite eq. (7) as $G(x,y) = \left(C_1 + \frac{C_2}{j+k}\right)\frac{\delta}{(u+n)^{a/2}}$.

It is noted that $k$ and $n$ are constants. Since $F(x,y)$ is convex with respect to $x = j + k$ and $y = u + n$, we know that $G(j,u)$ is convex with respect to $j$ and $u$.

Now, we use the first-order Taylor expansion to approximate the convex function $G(j,u)$ at any $j_0$ and $u_0$ as

$$G(j,u) \geq G(j_0, u_0) + G_j'(j_0, u_0)(j - j_0)$$
$$+ G_u'(j_0, u_0)(u - u_0) \tag{31.5}$$

where

$$G_j'(j_0, u_0) = \frac{-C_2\gamma}{(j_0 + k)^2 (u_0 + n)^{a/2}} \tag{31.6}$$

$$G_u'(j_0, u_0) = \frac{-(a/2)\gamma(C_1(j_0+k) + C_2)}{(j_0+k)(u_0+n)^{a/2+1}} \tag{31.7}$$

By letting $x_0 = 0, y_0 = 0$, we can obtain

$$\left(C_1 + \frac{C_2}{j+k}\right)\frac{\delta}{(u+n)^{a/2}} \geq \left(C_1 + \frac{C_2}{k}\right)\frac{\delta}{n^{a/2}} - \frac{C_2 \delta j}{k^2 n^{a/2}} - \frac{(\frac{a}{2})\delta(C_1 k + C_2)u}{kn^{a/2+1}} \tag{31.8}$$

Finally, we substitute $j = e^{-(B_1 + B_2 v_m[t])} - e^{-(B_1 + B_2 v_m^\mu[t])}$, $k = 1 + e^{-(B_1 + B_2 v_m^\mu[t])}$, $u = \|\mathbf{q}[t] - \mathbf{s}_m\|^2 - \|\mathbf{q}^\mu[t] - \mathbf{s}_m\|^2$, and $n = \|\mathbf{q}^\mu[t] - \mathbf{s}_m\|^2$ into inequality (31.8), and, then, get Lemma 4.

## APPENDIX F

*The method of finding the initial trajectory of the UAV.*

With any given task completion time $T$, we solve problem (16) and problem (26) iteratively to find the approximate solution to problem (15). The initial trajectory of the UAV, which is represented as $\mathbf{Q}^{\mu=0} = \{\mathbf{q}^{\mu=0}[t], \forall t \in \mathcal{T}\}$, is the initial point to perform the iterative algorithm. In this section, we implement an algorithm to determine $\mathbf{Q}^{\mu=0}$.

First, we assume that the UAV accesses each GT in turn at the maximum speed $v^{\text{max}}$. The objective of the path planning in this step is to minimize the total flight distance/time of the UAV. This problem is mapped to the traveling salesman problem (TSP), in which the UAV (as a salesman) starts from the data center, then visits each GT (as a city) and finally returns back to the data center. Although TSP is NP-hard, many methods, such as genetic algorithms have been developed to find the high-quality solution.

By solving the TSP, we obtain the minimum time $T^{\text{F}}$ for the UAV to access all the GTs. We assume that the $m^{\text{th}}$ GT is the $n$th GT to be visited by the UAV, and we represent the coordinate of the $m^{\text{th}}$ GT as $\mathbf{s}_m^n$. The waypoint of UAV to visit the GTs is given as $\mathbf{W} = (\mathbf{w}_0, \mathbf{w}_1, \cdots, \mathbf{w}_M, \mathbf{w}_{M+1})$, where $\mathbf{w}_0 = \mathbf{w}_{M+1} = \mathbf{q}^{\text{ini}}$, and $\mathbf{w}_n = (\mathbf{s}_m^n, H)$ for $\forall n = 1, 2, \ldots, M$.

With a given $T$, we derive the optimal waypoint of the UAV to access the GTs by analyzing the following two cases.
With a given $T$, we derive the optimal waypoint of the UAV to access the GTs by analyzing the following two cases.

1) If $T \geq T^{\text{F}}$, it means that the UAV is with enough time to fly right above each GT. Thus, the remaining $(T - T^{\text{F}})$ time slots can be evenly distributed among the $M$ GTs for the UAV to hover. In this case, the optimal waypoint of the UAV is $\mathbf{W}$.

2) If $T < T^{\text{F}}$, the given $T$ is not enough to make the UAV fly right above each GT. Similar to [4], we generate a circle with radius $r$ for each GT. The center of the circle is the position of the GT. With a given $T$, the waypoint of the UAV should fall in the circular area specified by each GT. In this case, the optimal waypoint of the UAV is achieved by solving the following problem.

$$\min_{\mathbf{W}} \frac{\sum_{n=1}^{|\mathcal{M}|+1} \|\mathbf{w}_n - \mathbf{w}_{n-1}\|}{v^{\text{max}}} \tag{32}$$

$$\text{s.t. } \|\mathbf{w}_n - \mathbf{w}_{n-1}\| \leq r, \forall n = 1, 2, \ldots, M. \tag{32.1}$$

Let $T^{\text{r}}$ denote the time for the UAV to access all the circular areas specified by the GTs. The optimal $T^{\text{r}}$ is obtained by solving problem (32). It is noted that $T^{\text{r}}$ decreases with the increasing $r$. Therefore, with any given $r$, we can get the optimal radius $r$ and the optimal waypoint $\mathbf{W}$ by using the BSA. Once the optimal waypoint **W** is obtained, we connect the points in **W**, and then get $(M+1)$ straight path segments. The initial trajectory of the UAV, $\mathbf{Q}^{\mu=0}$, is achieved by using linear interpolation.